\documentclass[11pt]{iopart}
\usepackage{graphicx}
\usepackage{epsf,amsfonts,amssymb,amsbsy,wrapfig,cite}
\begin{document}
\title[Radial geodesics as a microscopic origin of black hole entropy]
{Radial geodesics as a microscopic origin of black hole entropy.\\
II: Features of Reissner--Nordstr\o m black hole}
\author{V.V.Kiselev*\dag}%\ddag}
%\fax{}
\address{*%\dag
\ Russian State Research Center ``Institute for
High Energy
Physics'', %\\
Pobeda 1, Protvino, Moscow Region, 142281, Russia}
\address{\dag\ %\ddag\
Moscow
Institute of Physics and Technology, Institutskii per. 9,
Dolgoprudnyi Moscow Region, 141700, Russia}
%\date{}
\ead{kiselev@th1.ihep.su}
\begin{abstract}
The entropy of charged black hole is calculated by using the
partition function evaluated at radial geodesics confined under
horizons. We establish two quantum phase states inside the black
hole and a transition between them.
\end{abstract}
%\submitto{JCAP}
%\maketitle

%%%%%%%%%%%%%%%%%%%%%%%%%%%%%%%%%%%%%%%%%%%%%%%%%%%%%%%%%%%%%%%%%%

\section{Preface \textit{(instead of Introduction)}}

The second chapter of present paper is devoted to consequent
launching the constructed vehicle to a charged black hole. The
Reissner--Nordstr\o m black hole has got two horizons, that
constitutes a source of specific properties: We will show that the
thermal quantization leads to \textit{i)} quantizing a ratio of
horizon areas, \textit{ii)} two states of aggregation with
different temperatures and entropies, \textit{iii)} a possibility
of phase transition between two states. We derive the entropy of
Reissner-Nordstr\o m black hole in terms of partition function
evaluated by the action for causal radial geodesics confined under
the horizons. The interior structure of charged black hole
mathematically differs from that of Schwarzschild one: instead of
conic geometry with a single map on the manifold it has got a
sphere, which should be covered by two coordinated maps.

In section 2 we describe the mapping of causal radial geodesics
confined under the horizon and thermally quantize them. Section 3
is devoted to the evaluation of entropy. Results are summarized
in section 4.

\section{Radial geodesics}
\begin{figure}[th]
\centerline{\includegraphics[width=7cm]{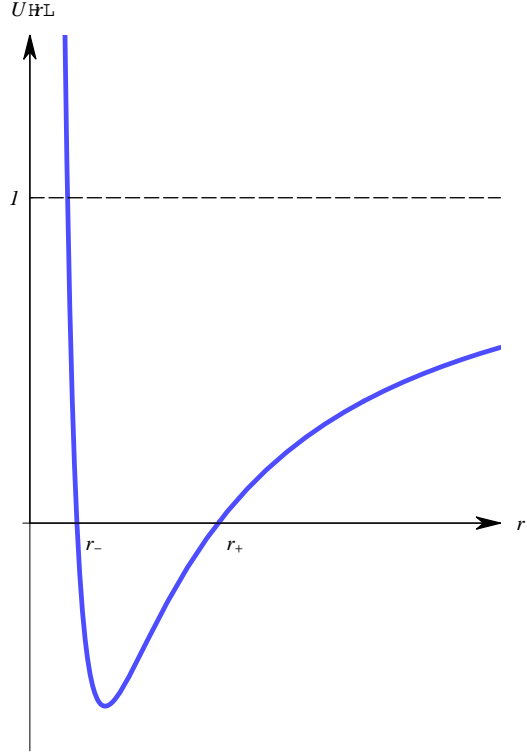}}
  \caption{The gravitational potential $U(r)$ for
  the Reissner--Nordstr\o m black hole.}\label{potential}
\end{figure}

The Reissner--Nordstr\o m  metric for a static spherically
symmetric charged black hole has the form
\begin{equation}\label{RN}
    {\rm d}s^2=g_{tt}(r)\,{\rm
    d}t^2-\frac{1}{g_{tt}(r)}\,{\rm d}r^2-r^2[{\rm
    d}\theta^2+\sin^2\theta\,{\rm d}\phi^2],
\end{equation}
with
\begin{equation}\label{gtt}
    g_{tt}(r) = \frac{1}{r^2}\,(r-r_+)(r-r_-),
\end{equation}
whereas the horizon points determine a black hole mass $M$ and its
charge $Q$ by
\begin{equation}\label{M-Q}
    M=\frac{1}{2}(r_++r_-),\qquad Q^2= r_+ r_-,
\end{equation}
in units $G=c=\hbar=1$. In a standard manner, an introduction of
\begin{equation}\label{r_*}
    r_*=\int\frac{{\rm d}r}{g_{tt}(r)}
\end{equation}
transforms the metric of (\ref{RN}) to the form
\begin{equation}\label{RN*}
    {\rm d}s^2=g_{tt}(r)\,({\rm
    d}t^2-{\rm d}r_*^2)-r^2[{\rm
    d}\theta^2+\sin^2\theta\,{\rm d}\phi^2],
\end{equation}
where
\begin{equation}\label{r_*2}
    r_* %%=\int\frac{r^2\,{\rm d}r}{(r-r_+)(r-r_-)}
    =r+\frac{r_+^2}{r_+-r_-}\,\ln\left[\frac{r}{r_+}-1\right]-
    \frac{r_-^2}{r_+-r_-}\,\ln\left[\frac{r}{r_-}-1\right].
\end{equation}
Following the Hamilton--Jacobi method for radial geodesics of a
particle with mass $m$, we get
\begin{equation}\label{dot r_*}
    \left(\frac{{\rm d}r_*}{{\rm d}t}\right)^2=1-A\,g_{tt}(r),
\end{equation}
where $A=m^2/{\cal E}^2$ is the only integral of radial motion, so
that the Hamilton--Jacobi action $S_{HJ}$
\begin{equation*}
    S_{HJ}=-{\cal E}\,t+{\cal S}_{HJ}(r_*)
\end{equation*}
satisfies the equation
\begin{equation}\label{HJ}
    \frac{1}{m^2}\left(\frac{\partial {\cal S}_{HJ}}{\partial
    r_*}\right)={\cal E}_A- U(r),
\end{equation}
with a scaled energy ${\cal E}_A=1/A$, and the potential
$U(r)=g_{tt}(r)$ as shown in Fig. \ref{potential}.

For a massive particle, the interval on a radial geodesic curve is
given by
\begin{equation}\label{dsRN}
    {\rm d}s^2 =\frac{A\,{\rm d}r^2}{1-A\,g_{tt}(r)},
\end{equation}
and we call it `causal' if it is time-like
\begin{equation*}
    {\rm d}s^2 > 0,
\end{equation*}
which takes place at
\begin{equation}\label{positive}
    \{A >0 \}\;\cup\;\left\{A < -\frac{4 r_+ r_-}{(r_+
    -r_-)^2}<0\right\},
\end{equation}
and
\begin{equation}\label{roots}
   \{ r_1 < r < r_2, \, A>1\cup A<0 \}\,\cup\,\{ r_1<r\leqslant+\infty,\,0< A<1\}
\end{equation}
where the return points have to be given by real roots of
quadratic equation
\begin{equation*}
    (1-A)\,r^2+A(r_++r_-)\,r-A\,r_+r_-=0
\end{equation*}
with discriminant \begin{equation*}
    {\cal D} = A[A(r_+ -r_-)^2+4 r_+ r_-] \geqslant 0.
\end{equation*}
For geodesics confined under the horizons $A<-{4 r_+ r_-}/{(r_+
-r_-)^2}$, the roots are arranged as
\begin{equation*}
    r_- \leqslant r_1\leqslant r_2 \leqslant r_+.
\end{equation*}
Particles at such the geodesics compose a thermal ensemble, since
the condition of causality in terms of time
\begin{equation}\label{>0}
    {\rm d}s^2 = A\,g^2_{tt}(r)\,{\rm d}t^2 > 0
\end{equation}
implies the imaginary ${\rm d}t$.

Kruskal coordinates are introduced by
\begin{equation}\label{i1}
\left\{\begin{array}{l}
 u=t-r_*,\\[1mm]
 v=t+r_*,\end{array}
 \right.
\end{equation}
and two maps marked by ``$+$'' and ``$-$''
\begin{equation}\label{i2+}
    \left\{\begin{array}{l}
%\vspace*{-10mm}\\
\displaystyle\bar u_+=-\frac{2 r_+^2}{r_+ -r_-}\,
\exp\left[{\displaystyle-\frac{u}{2r_+^2}(r_+ -r_-)}\right]
%= -2r_g\,e^{\displaystyle\frac{r-t}{2r_g}}\,\sqrt{%\left(
%\frac{r}{r_g}-1%\right)}
,\\[4mm]
\displaystyle\bar v_+=+\frac{2 r_+^2}{r_+ -r_-}\,
\exp\left[{\displaystyle+\frac{v}{2r_+^2}(r_+ -r_-)}\right]
%= +2r_g\,e^{\displaystyle\frac{r+t}{2r_g}}\,\sqrt{%\left(
%\frac{r}{r_g}-1%\right)}\\[1mm]
,%%\\[3mm]
\end{array}
 \right. \quad \mbox{at } r_-<r<\infty,
\end{equation}
\begin{equation}\label{i2-}
    \left\{\begin{array}{l}
%\vspace*{-10mm}\\
\displaystyle\bar u_-=+\frac{2 r_-^2}{r_+ -r_-}\,
\exp\left[{\displaystyle+\frac{u}{2r_-^2}(r_+ -r_-)}\right]
%= -2r_g\,e^{\displaystyle\frac{r-t}{2r_g}}\,\sqrt{%\left(
%\frac{r}{r_g}-1%\right)}
C_-,\\[4mm]
\displaystyle\bar v_-=-\frac{2 r_-^2}{r_+ -r_-}\,
\exp\left[{\displaystyle-\frac{v}{2r_-^2}(r_+ -r_-)}\right]
%= +2r_g\,e^{\displaystyle\frac{r+t}{2r_g}}\,\sqrt{%\left(
%\frac{r}{r_g}-1%\right)}\\[1mm]
C_+,%%\\[3mm]
\end{array}
 \right. \quad \mbox{at } 0<r<r_+,
\end{equation}
where $C_\pm$ compensate complex phases of $r_*$ at $r_-<r<r_+$,
for instance: $C_\pm=\exp[{\rm i}\pi (r_+^2/r_-^2+l)]$, $l\in
\mathbb N$.

\begin{figure}[th]
  \centerline{\includegraphics[width=5cm]{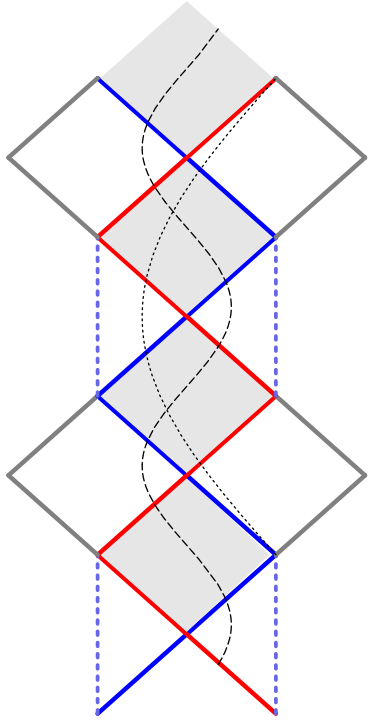}}
  \caption{The Reissner--Nordstr\o m space-time in compactified variables.
  The singularity $r=0$ is shown by bold dashes. The horizons are borders of
  black hole interior marked by shaded polygons. A short-dashed curve represents
  a geodesic line reaching the infinity, while a long-dashed curve does a geodesic line
  oscillating between maximal and minimal radii. Mirror geodesics are given
  by reflection with respect to the vertical axis. The top and bottom of
  tape should be glued (taking into account that the geodesics will not return
  to regions, wherein they visited in past).}
  \label{RN-spacetime}
\end{figure}

One could compactify the above representations by substitution
$w\to 2\arctan [w]$ for each variable to get a scheme of causal
structure \cite{BirrelDavies} for the Reissner--Nordstr\o m
space-time shown in Fig. \ref{RN-spacetime}.

Two maps give regular metric near the horizons, but it has the
true singularity at $r=0$ and a coordinate singularity at
$r=\infty$, which is removed by mapping in the $\{t,r\}$ plane.

\begin{figure}[th]
  \centerline{\includegraphics[width=11cm]{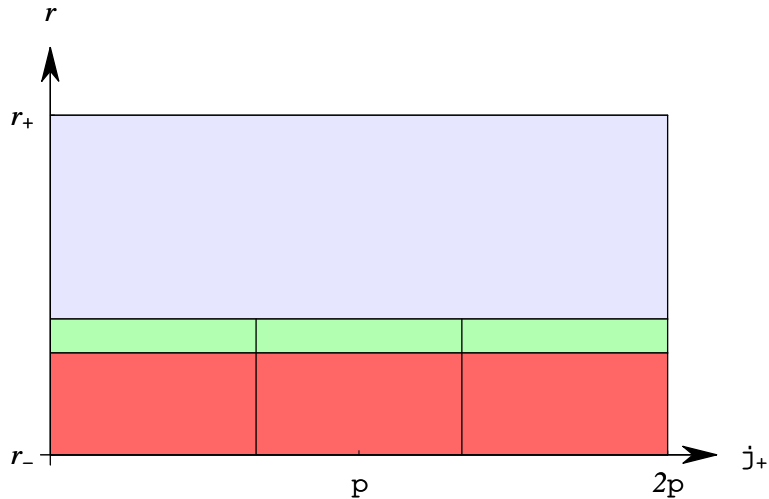}}
  \caption{The mapping of interior;
  the phases of aggregation in the charged black hole: a cool ``ice''
  (top), a hot ``water'' (bottom), and the transition layer of ``melting
  ice'' (middle).}\label{phases}
\end{figure}

Appropriate coordinates for geodesics completely confined under
the horizons are given by
\begin{equation}\label{eucliddef+}
    \left\{\begin{array}{l}
 \bar u_+=\,\varkappa_+\,{\rm i}\,\rho_+\,e^{{\rm i}\,\varphi_+},\\[2mm]
 \displaystyle
 \bar v_+=-\frac{{\rm i}}{\varkappa_+}\,\rho_+\,e^{-{\rm i}\,\varphi_+},\end{array}
 \right.
 \qquad
 \left\{\begin{array}{l}
 \bar u_-=\,\varkappa_-\,{\rm i}\,\rho_-\,e^{{\rm i}\,\varphi_-},\\[2mm]
 \displaystyle
 \bar v_-=-\frac{{\rm i}}{\varkappa_-}\,\rho_-\,e^{-{\rm i}\,\varphi_-},\end{array}
 \right.
\end{equation}
or (at `initial' times $\Delta t_0^\pm =- \frac{2r^2_\pm}{r_+
-r_-}\ln\varkappa_\pm$, which are set to zero)
\begin{equation}\label{euclid-def2+}
    \left\{\begin{array}{l}
    \displaystyle
\;\, t=\frac{r^2_\pm}{r_+ -r_-}\,\ln e^{-2{\rm
i}\,\varphi_\pm}=-{\rm
 i}\,\frac{2r^2_\pm}{r_+
-r_-}\,\varphi_\pm,\qquad \varphi_+\in [0,2\pi],
 \\[4mm]\displaystyle
 r_*= \frac{2r^2_\pm}{r_+
-r_-} \ln\left[-\frac{\rho_\pm}{2 r^2_\pm}(r_+ -r_-)\right],
  \hfill\rho_\pm \in [0,\infty],\end{array}
 \right.
\end{equation}
while two maps are consistent with each to other if only the ratio
of periods in $\varphi_\pm$ is a natural number $l$, so that the
periods in $t$ are also consistent,
\begin{equation}\label{quant-r}
    \frac{r_+^2}{r_-^2}=l \in \mathbb N.
\end{equation}
Remember that the periods in the imaginary time give the inverse
temperature $\beta$ of thermal ensemble. Therefore, we deduce that
horizons are in general at different temperatures, which ratio is
quantized:
\begin{equation}\label{temp}
    \beta_+ =\frac{4\pi r^2_+}{r_+ -r_-} >
    \beta_- =\frac{4\pi r^2_-}{r_+ -r_-},\qquad\qquad
%\end{equation}
%\begin{equation}\label{quant-t}
    \frac{\beta_+}{\beta_-} =l.
\end{equation}
An example of consistent mapping at $l=3$ is shown in Fig.
\ref{phases}, where we use the convenient plane of
$\{\varphi_+,r\}$.

\begin{figure}[th]
  \centerline{\includegraphics[width=8cm]{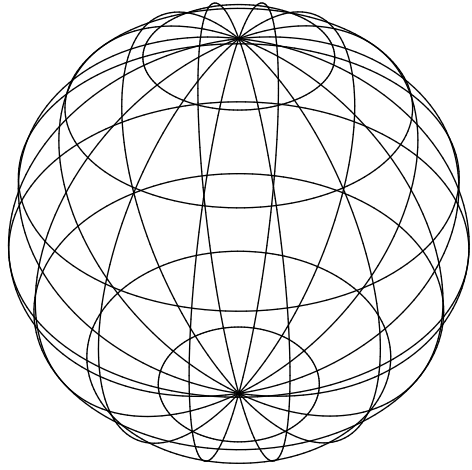}}

  \vspace*{-1cm}
  \caption{The compactified hole with poles at $r=r_\pm$.}
  \label{sphere}
\end{figure}

Since $\rho_\pm\to 0$ at $r\to r_\pm$ correspondingly, the
interior space-time of confined geodesics is a sphere with poles
at $r=r_\pm$, wherein the horizons are contracted, as pictured in
Fig. \ref{sphere}.

\begin{figure}[th]
\setlength{\unitlength}{1mm}
\begin{center}
\begin{picture}(146,20)
 \put(-4,0){\includegraphics[width=7cm]{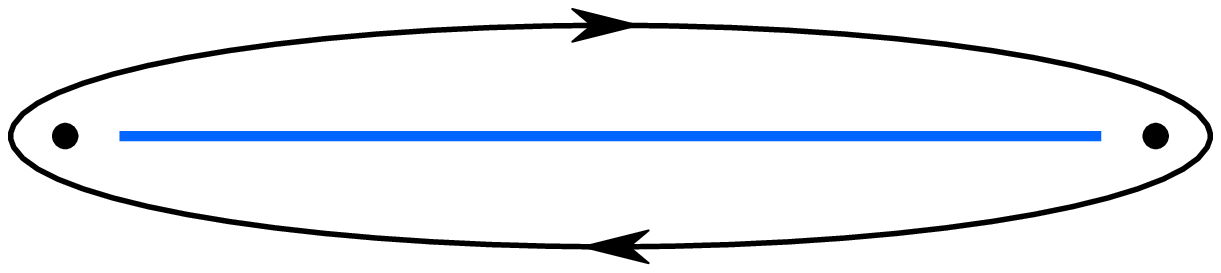}}
 \put(69,6){$\equiv$}
 \put(76,0){\includegraphics[width=7cm]{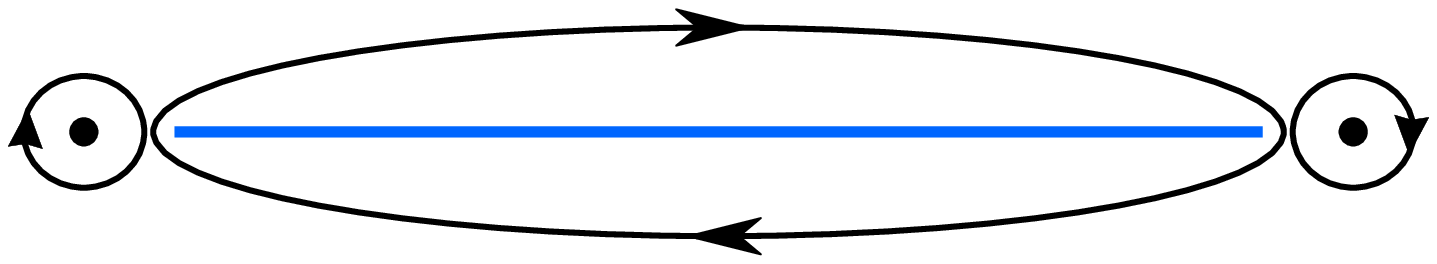}}
\end{picture}
\end{center}
\caption{The integration contour. The cut connects peculiar points
of square root in (\ref{Dt}), while dots denote the poles of
inverse $g_{tt}(r)$.} \label{contour}
\end{figure}

The ground state of ${\cal E}_A\to 0$ is easily calculable. So,
the integral increment of interval per cycle is given by
\begin{equation}\label{Ds}
    \Delta_c s = 2\int\limits_{r_-}^{r_+}\frac{r{\rm
    d}r}{\sqrt{(r_+-r)(r-r_-)}}= \pi\,(r_++r_-),
\end{equation}
while the time increment
\begin{equation}\label{Dt}
    \Delta_c t_E =\int\limits_{\Delta_c}\frac{{\rm d}t_E}{{\rm
    d}r}\,{\rm d}r =-{\rm i}\int\limits_{\Delta_c}\frac{{\rm
    d}r}{g_{tt}(r)}\,\frac{1}{\sqrt{1-A\,g_{tt}(r)}}
\end{equation}
can be simply evaluated at $A\to -\infty$, since in this limit the
integral over the contour shown in Fig. \ref{contour} should be
equal to zero, because the integrand does not cross any
singularity, and it tends to zero.

\noindent
 Therefore, the increment is given by the residuals in the poles
 of inverse $g_{tt}(r)$,
 \begin{equation}\label{Dt2}
    \Delta_c t_E = -{\rm i}\,2\pi{\rm i}\left[\frac{1}{g^\prime_{tt}(r_+)}
    +\frac{1}{g^\prime_{tt}(r_-)}\right] =2\pi\,(r_++r_-).
\end{equation}
Winding numbers are given by
\begin{equation}\label{n}
    n_\pm=\frac{2\pi}{\Delta_c\varphi_\pm}=\frac{4\pi r_\pm^2}{(r_+-r_-)\Delta_c
    t_E},
\end{equation}
so that
\begin{equation}\label{w}
    n_+=\frac{2 l}{l-1} \in \mathbb N, \qquad n_-=\frac{2}{l-1} \in \mathbb
    N.
\end{equation}
Then, we can list quantized winding numbers as given in Table
\ref{wind}.

\begin{table}[th]
  \centering
  \caption{Admissible values of winding numbers $n_\pm$ in the ground state
  versus the quantum ratio of
  horizon areas $l$.}\label{wind}
  \begin{tabular}{|c|c|c|}
    \hline
    $l$ & $n_+$ & $n_-$ \\
    \hline
    $1$ & $\infty$ & $\infty$\\
    $2$ & $4$ & $2$\\
    $3$ & $3$ & $1$\\
    $\infty$ & $2$ & $0$\\
    \hline
  \end{tabular}
\end{table}

%\noindent
 The limit cases of $l=1$ and $l=\infty$ correspondingly give an extremal charged
black hole ($r_+=r_-$) and Schwarzschild one (uncharged). There
are four kinds of quantum charged black holes in the ground state,
only.

Generically, at admissible $A$ under horizons we get
\begin{equation}\label{DstA}
    \Delta_c^A s =x^{3/2}\cdot\Delta_c s,
\end{equation}
where again
\begin{equation}\label{x}
    x=\frac{-A}{1-A},
\end{equation}
while the winding numbers are given by
\begin{equation}\label{windA}
    n_\pm^A= \frac{2\pi}{\Delta_c^A\varphi_\pm},
\end{equation}
with
\begin{equation}\label{pA}
    \Delta_c^A\varphi_\pm =
    \frac{\pi}{2}\,\frac{r^2_+
    -r^2_-}{r^2_\pm}\left[2-(x+2)\sqrt{1-x}\right].
\end{equation}
The action of particle is determined by
\begin{equation}\label{act}
    S_n =-mc\cdot n^A_\pm\, \Delta_c^A
    s=-mc\cdot\beta_\pm\,\frac{x^{3/2}}{2-(x+2)\sqrt{1-x}},
\end{equation}
which is the same function of quantized $x$ as was studied in the
case of Schwarzschild black hole. The value at the ground level
($x\to 1$) is equal to
\begin{equation}\label{ground}
    S_{\rm gr} =-mc\cdot \frac{\beta_\pm}{2}.
\end{equation}

\section{Entropy}
Following the method invented in the case of Schwarzschild black
hole, we evaluate the partition function by making use of the
quantum action gap:
\begin{equation}\label{sum}
    \ln Z\approx \sum S_{\rm gr} = -\frac{\beta_+}{2}\,\sigma_+-
    \frac{\beta_-}{2}\,\sigma_-,
\end{equation}
where we have introduced the sums of masses on geodesics with
corresponding maps,
\begin{equation}\label{masses}
    \sigma_\pm =\sum_\pm mc.
\end{equation}
For the sake of simplicity and spectacular clarity, let us, first,
evaluate the partition function for a pure ``$+$''-state
($\sigma_-=0$), that has a temperature less then a pure
``$-$''-state ($\sigma_+=0$), which we respectively call ``ice``
and ``water''. Then, the solution for $\sigma$ is explicit:
\begin{equation}\label{s+}
    \sigma_+ =2M-\frac{1}{2\beta_+}\,{\cal A_+},
\end{equation}
where ${\cal A_+} =4\pi\,r_+^2$ is the area of horizon at $r=r_+$.
Indeed, since the temperature $T_+=1/\beta_+$ is determined by a
`surface gravity' \cite{BirrelDavies,Hawking2} as
\begin{equation}\label{T}
    T_+ =4\,\frac{\partial M}{\partial{\cal A}_+},\qquad \mbox{at }{\rm
    d}Q^2\equiv 0,
\end{equation}
for the average energy equal to the mass we get the following
identity:
\begin{equation}\label{M}
\begin{array}{rcl}
    M &=&\displaystyle
    \langle E\rangle =-\frac{\partial \ln Z_+}{\partial \beta_+}=
    \frac{\partial}{\partial\beta_+}(\beta_+ \sigma_+)/2\\[5mm]
    &=&\displaystyle
    M-\frac{1}{4\beta_+}\,{\cal A}_+
    +\frac{\beta_+}{2}\left[2 \frac{\partial M}{\partial\beta_+}+\frac{1}{2\beta_+^2}\,
    {\cal A}_+-\frac{1}{2\beta_+}\,\frac{\partial{\cal
    A}_+}{\partial\beta_+}\right]=M.
\end{array}
\end{equation}
Then, the entropy of ``ice'' is equal to
\begin{equation}\label{S+}
    {\cal S}_+ =\frac{1}{4}\,{\cal A}_+\,.
\end{equation}

We can repeat the same manipulations with the pure ``water'' state
of aggregation\footnote{Formally, the Christodoulou--Ruffini
formula $M^2=[(Q^2+{\cal A}/4\pi)^2+4J^2] \pi/{\cal A}$, yielding
the black hole mass in terms of (outer/inner) horizon area ${\cal
A}={\cal A}_\pm$, charge $Q$ and angular momentum $J$, gives
negative value of temperature at the inner horizon
$T_-=-1/\beta_-$, which corresponds to the black body Hawking
radiation confined in the inner space-time at $r<r_-$. The
situation is analogous to inverse saturation of levels in lasers.
Geometrically, one could also relate the negative sign with the
opposite orientation of inner horizon surface with respect to the
outer one.} and get its entropy
\begin{equation}\label{S-}
    {\cal S}_- =\frac{1}{4}\,{\cal A}_-\,.
\end{equation}
Thus, the ``water'' has the entropy less than the ``ice'', as well
as the greater temperature, that means the layer of ``melting
ice'' between two phases is not stable: the interior is cooling to
the ``ice''; the quantum transition of aggregation states takes
place with a jump of temperature and entropy. We have just found
that the space-time outside the external horizon forms the cool
thermostats, while the space-time inside the inner horizon forms
the hot thermostats. For the extremal black hole both thermostats
have the identical temperature (the hole does not radiate), while
the temperature of singularity in the Schwarzschild black hole is
equal to infinity (the hole radiates, since it is heated by the
singularity). To our opinion, thermodynamically the singularity
should cool off, as well as the inner horizon does.

As for a mixed stage of ``ice'' and ``water'', one could introduce
fractions of aggregate states, $f_\pm$, so that
\begin{equation}\label{frac}
    \ln Z =f_+\ln Z_++f_-\ln Z_-,\qquad f_++f_-=1,
\end{equation}
and
\begin{equation*}
    M=-f_+\,\frac{\partial \ln Z_+}{\partial \beta_+}
    -f_-\,\frac{\partial \ln Z_-}{\partial \beta_-}\equiv M.
\end{equation*}

Thus, the Bekenstein--Hawking entropy \cite{Bekenstein,Hawking}
gives the entropy of pure ``ice''.

As we have already seen in chapter I, the entropy evaluated by the
partition function represents the number of microstates with
various winding numbers.

\section{Discussion and conclusion}

In present chapter we have expanded the method based on the
consideration of radial geodesics confined under the horizons, to
the charged black hole. We have found the quantization of areas of
horizons due to the consistency of maps covering the black hole
interior. Two states of aggregation have been deduced: the cool
``ice'' and hot ``water'', which can be exposed by the quantum
phase transition.

The ground state coherently matches all points $r$ inside the
interior of black hole to its external surface at $r_+$.
Therefore,  we do not expect any straightforward contradiction
with estimates based on the Cardy formula \cite{Cardy} in
conformal field theories
 \cite{Carlip,Carlip2,Strominger,StromingerVafa+ },
 in agreement with the principle of holography \cite{Holograph}.
 The same note concerns for an evaluation of quantum field information eaten
 by the black hole \cite{FrolovNovikov}. Note, that the
 extremal black hole is described by solitonic BPS states in
 superstrings \cite{StromingerVafa+}. The most general case of
 extremal black hole involves its rotation. So, we are permanently
 travelling to an analogous exercise with the
Kerr-Newman black hole (a charged and rotating one), which will be
considered in chapter III.

\vspace*{3mm}
 This work is partially supported by the grant of the
president of Russian Federation for scientific schools
NSc-1303.2003.2, and the Russian Foundation for Basic Research,
grants 04-02-17530, 05-02-16315.

%\newpage

\end{document}